\documentclass[aps,prb,preprint,showpacs,showkeys,superscriptaddress]{revtex4-1}

\usepackage{amsmath}
\usepackage{graphics}

\include{MyMc}
\include{we}

\begin{document}

\title{Josephson junction with magnetic-field tunable current-phase relation}

\author{A. Lipman}
\affiliation{%
  The Raymond and Beverly Sackler School of Physics and Astronomy,
  Tel Aviv University, Tel Aviv 69978, Israel
}

\author{R. G. Mints}
\affiliation{%
  The Raymond and Beverly Sackler School of Physics and Astronomy,
  Tel Aviv University, Tel Aviv 69978, Israel
}

\author{R. Kleiner}
\author{D. Koelle}
\author{E. Goldobin}
\affiliation{%
  Physikalisches Institut and Center for Collective Quantum Phenomena in LISA$^+$,
  Universit\"at T\"ubingen, Auf der Morgenstelle 14, D-72076 T\"ubingen, Germany
}

\date{%
  \today\ 
}

\begin{abstract}
  We consider a 0-$\pi$ Josephson junction consisting of asymmetric $0$ and $\pi$ regions of different lengths $L_0$ and $L_\pi$ having different critical current densities $j_{c,0}$ and $j_{c,\pi}$. If both segments are rather short, the whole junction can be described by an \emph{effective} current-phase relation for the spatially averaged phase $\psi$, which includes the usual term $\propto\sin(\psi)$, a \emph{negative} second harmonic term $\propto\sin(2\psi)$ as well as the unusual term $\propto H \cos\psi$ tunable by magnetic field $H$. Thus one obtains an electronically tunable current-phase relation. At $H=0$ this corresponds to the $\varphi$ Josephson junction. 

\end{abstract}



\maketitle

\section{Introduction}
\label{Sec:Intro}

Recently we proposed\cite{Goldobin:2011:0-pi:H-tunable-CPR} to implement a $\varphi$ Josephson junction (JJ) \cite{Buzdin:2003:phi-LJJ} with magnetic-field tunable current-phase relation (CPR) based on an 0-$\pi$ JJ with the 0 and $\pi$ segments of different length $L_0\neq L_\pi$. This proposal was made keeping in mind \YBCO-Nb ramp zigzag JJ technology\cite{Smilde:ZigzagPRL,Hilgenkamp:zigzag:SF} (or a similar one\cite{Ariando:Zigzag:NCCO} with \NCCO-Nb) established recently in our group also\cite{Scharinger:ramp-zigzag:Ic(H)}. However, in experiment we were more successful\cite{Sickinger:varphiExp} in employing superconductor-insulator-ferromagnet-superconductor (SIFS) 0-$\pi$ JJs\cite{Weides:2006:SIFS-0-pi,Weides:2010:SIFS-jc1jc2:Ic(H),Kemmler:2010:SIFS-0-pi:Ic(H)-asymm}, where the lengths of 0 and $\pi$ segments are equal, but critical current densities $j_{c,0}$ and $j_{c,\pi}$ in the 0 and $\pi$ parts are different. 

Therefore, in this paper we present a more general theory, which describes an effective $\varphi$ JJ made of asymmetric $0$ and $\pi$ regions of different lengths $L_0$ and $L_\pi$ having different critical current densities $j_{c,0}$ and $j_{c,\pi}$.

\section{Model}
\label{Sec:Model}

The static sine-Gordon equation that describes the behavior of the Josephson phase $\phi$ in a 0-$\pi$ JJ is
\begin{equation}
  \frac{\Phi_0}{2\pi\mu_0 d_J}\phi'' - j_c(x)\sin\phi = -j
  .\label{Eq:sine-Gordon:Phys}
\end{equation}
Here $\mu_0$ is the magnetic flux quantum, $\mu_0 d_J$ is the specific inductance (per square) of the superconducting electrodes forming the JJ and $j$ is the bias current density. The prime denotes the partial derivatives with respect to coordinate $x$. We assume that the critical current density $j_c(x)$ has the form of a step-function
\begin{eqnarray}
  &j_c=j_{c,0}>0,\quad\ &0\le x \le L_0
  ,\label{Eq:eq01}\\
  \label{eq02}
  &j_c=j_{c,\pi}<0,\quad\ &-L_\pi\le x < 0.
\end{eqnarray}
We write the critical current density $j_c(x)$ as
\begin{equation}
  j_c(x)=\av{j_c(x)} [1+g(x)]
  , \label{eq04}
\end{equation}
where
\begin{equation}
  \av{j_c} = \frac{1}{L}\int_{-L_\pi}^{L_0} j_c(x)\,dx
  = \frac{1}{L}\left( j_{c,0}L_0 + j_{c,\pi} L_\pi \right)
  \label{Eq:j_av}
\end{equation}
is the average critical current density, $L=L_0+L_\pi$ is the total length of the junction, and
$\av{g(x)}=0$. The function $g(x)$ is defined as
\begin{equation}\label{eq07}
  g(x) = \frac{j_c(x)}{\av{j_c(x)}} - 1
\end{equation}
that results in
\begin{equation}
  g(x) =
  \begin{cases}
    g_0,   &      0<x<L_0,\\
    g_\pi, & -L_\pi<x<0.
  \end{cases}
  \label{Eq:g(x)}
\end{equation}
where
\begin{equation}
  g_0   = \frac{(j_{c,0}-j_{c,\pi})L_\pi}{j_{c,0}L_0+j_{c,\pi} L_\pi}
  ;\quad
  g_\pi =-\frac{(j_{c,0}-j_{c,\pi})L_0  }{j_{c,0}L_0+j_{c,\pi} L_\pi}
  . \label{Eq:g}
\end{equation}

Then we divide Eq.~\eqref{Eq:sine-Gordon:Phys} by $|\av{j_c}|$ and normalize the coordinate $x$ to the Josephson length calculated using $|\av{j_c}|$, \ie,
\begin{equation}
  \lambda_J = \sqrt{\frac{\Phi_0}{2\pi\mu_0 d_J |\av{j_c}|}}
  . \label{Eq:lambda_J_av}
\end{equation}
Thus, we obtain a normalized sine-Gordon equation for the phase difference $\phi(x)$
\begin{equation}
  \phi'' -\sgn(\av{j_c})\left[1+g(x)\right]\sin\phi =-\gamma
  , \label{Eq:sG(g):sgn}
\end{equation}
where $\gamma =j/|\av{j_c}|$ is the normalized bias current density. It is worth mentioning that $\av{j_c}$ can be positive as well as negative. Below, for the same of simplicity, we assume $\av{j_c}>0$. Thus, Eq.~\eqref{Eq:sG(g):sgn} becomes
\begin{equation}
  \phi'' -\left[1+g(x)\right]\sin\phi =-\gamma
  . \label{Eq:sG(g)}
\end{equation}
In the case $\av{j_c}<0$ the substitution $\phi\to\pi-\phi$ converts Eq.~\eqref{Eq:sG(g):sgn} to the same Eq.~\eqref{Eq:sG(g)}.

We look for a solution of Eq.~\eqref{Eq:sG(g)} in the form
\begin{equation} \label{Eq:PhaseSplit}
  \phi(x) = \psi + \xi(x)\sin\psi,
\end{equation}
where
\begin{equation}
  \psi = \av{\phi(x)}
  \label{Eq:Def:psi}
\end{equation}
is a constant \emph{average phase}, while $\xi(x)\sin\psi$ describes the deviation of the phase from the average value, \ie, $\av{\xi(x)}=0$. Further we assume that the deviation is small, \ie,  $| \xi(x)\sin\psi| \ll 1$. Then we plug the relation \eqref{Eq:PhaseSplit} into Eq.~\eqref{Eq:sG(g)}, expand it in series in $\xi(x)\sin\psi$, and keep the terms of zero and first order. We get
\begin{equation}
  \xi''\sin\psi - [1+g(x)][1+\xi(x)\cos\psi]\sin\psi = -\gamma.
  \label{Eq:sG-mix}
\end{equation}
The constant terms (zero order of $\xi$ in Eq.~\eqref{Eq:sG-mix}) are
\begin{equation}
  \gamma = \sin\psi + \av{\xi(x)g(x)}\cos\psi\sin\psi.
  \label{Eq:const-terms}
\end{equation}
The terms of first order of $\xi(x)$ in Eq.~\eqref{Eq:sG-mix} are
\begin{equation}
  \xi'' - g(x) = \{\xi + \xi(x)g(x)-\av{\xi(x)g(x)}\}\cos\psi.
  \label{Eq:dev-terms}
\end{equation}

Numerical calculations show that the two terms $\propto\cos\psi$ have an extremely weak effect on solutions of Eq.~\eqref{Eq:dev-terms}. We neglect these terms and obtain for $\xi(x)$
\begin{equation}
  \xi'' - g(x) = 0
  .\label{Eq:for-xi}
\end{equation}
We treat solutions of Eq.~\eqref{Eq:for-xi} by using the matching continuity (at $x=0$) and boundary (at $x=-l_\pi\equiv -L_\pi/\lambda_J$, $x=l_0\equiv L_0/\lambda_J$) conditions
\begin{equation}\label{eq16}
  \xi_{\pi}(0) = \xi_0(0),\qquad \xi_{\pi}'(0) = \xi_0'(0),
\end{equation}
\begin{equation}\label{eq17}
  \xi_\pi'(-l_\pi)\sin\psi = h,\qquad \xi_0'(l_0)\sin\psi = h.
\end{equation}

The applied field $H$ is normalized by $H_{c1}/2$, \ie,
\begin{equation} \label{eq18}
  h=\frac{2H}{H_{c1}},\qquad H_{c1}=\frac{\Phi_0}{\pi\Lambda\lambda_J},
\end{equation}
where $\Lambda$ is the effective magnetic thickness of the JJ. We integrate Eq. (\ref{Eq:for-xi}) once and obtain
\begin{eqnarray}\label{eq19}
  \xi_0'(x) = g_0(x-l_0) + \frac{h}{\sin\psi},\quad 0<x<l_0,
\end{eqnarray}
\begin{eqnarray}\label{eq20}
  \xi_{\pi}'(x) = g_{\pi} (x+l_\pi) + \frac{h}{\sin\psi},\quad -l_\pi<x<0.
\end{eqnarray}
The second integration results in
\begin{eqnarray}\label{eq21}
  \xi_0(x) = g_0 \left(\frac{x^2}{2}-l_0 x\right) + \frac{hx}{\sin\psi}+C,\\
  \text{for }0<x<l_0,\nonumber
\end{eqnarray}
\begin{eqnarray}\label{eq22}
  \xi_{\pi}(x) = g_{\pi}\left(\frac{x^2}{2}+l_\pi x\right) + \frac{hx}{\sin\psi}+C,\\
  \text{for }-l_\pi<x<0.\nonumber
\end{eqnarray}
The integration constant $C$ can be obtained using the condition $\langle\xi(x)\rangle = 0$
\begin{equation}\label{eq23}
  C=\frac{l_0 -l_\pi}{2}\left(\frac{g_0l_0+g_\pi l_\pi}{3}-\frac{h}{\sin\psi}\right).
\end{equation}

We use  Eqs.~\eqref{Eq:g(x)}, \eqref{eq21}, and \eqref{eq22} and obtain the average $\av{\xi(x)g(x)}$ in the form
\begin{equation}\label{Eq:xig}
  \av{\xi(x)g(x)} = \Gamma_0 + \Gamma_h\frac{h}{\sin\psi},
\end{equation}
where the coefficients $\Gamma_0$ and $\Gamma_h$ are given by
\begin{eqnarray}
    \Gamma_0 &=& -\frac{l_0^2l_\pi^2}{3} \frac{(j_{c,0}-j_{c,\pi})^2}{(j_{c,0}l_0+j_{c,\pi} l_\pi)^2}
    ,\label{Eq:Gamma0}\\
    \Gamma_h &=& \frac{l_0l_\pi}{2} \frac{j_{c,0}-j_{c,\pi}}{j_{c,0}l_0+j_{c,\pi} l_\pi}
    .\label{Eq:GammaH}
  \label{Eq:Gammas}
\end{eqnarray}

Using Eqs.~\eqref{Eq:const-terms} and \eqref{Eq:xig} we find the current-phase relation in the form
\begin{equation}
  j=\av{j_c}\left(\sin\psi +\Gamma_0\sin\psi\cos\psi +h\Gamma_h\cos\psi\right)
  .\label{Eq:CPR}
\end{equation}

It is worth noting that there is a simple relation between the coefficients $\Gamma_0$ and $\Gamma_h$. Indeed, it follows from Eqs. (\ref{Eq:Gamma0}) and (\ref{Eq:GammaH}) that
\begin{equation}
  \Gamma_0 = -\frac{4}{3}\Gamma_h^2
  .\label{Eq:GammaH(Gamma0)}
\end{equation}

In the case of equal lengths of 0 amd $\pi$ parts ($l_0=l_\pi =l/2$) we find
\begin{equation}
  \Gamma_0=-\frac{l^2}{12} \left(\frac{j_{c,0}-j_{c,\pi}}{j_{c,0}+j_{c,\pi}}\right)^2 , \qquad
  \Gamma_h= \frac{l}{4} \frac{j_{c,0}-j_{c,\pi}}{j_{c,0}+j_{c,\pi}}\,.
  \label{Eq:Gammas@EqualL}
\end{equation}

The energy $U(\psi)$ corresponding to the current-phase relation (\ref{Eq:CPR}) is given by
\begin{equation}
  U(\psi)=\av{j_c}\left(1-\cos\psi +h\Gamma_h\sin\psi+\frac{\Gamma_0}{2}\sin^2\psi\right)
  . \label{Eq:U(psi)}
\end{equation}

\section{Conclusions}
\label{Sec:Conclusions}

\iffalse
Actually we've made a lot of stuff: extended theory for $|\xi(x)\sin\psi|\ll1$ to the case of $L_0\neq L_\pi$ and $j_{c,0}\neq|j_{c,\pi}|$; wrote solution for semifluxon in infinite asymmetric LJJ, ...
\else
We have extended our previous results \cite{Goldobin:2011:0-pi:H-tunable-CPR} to the case of arbitrary critical current densities $j_{c,0}\neq j_{c,\pi}$ more relevant for experiment\cite{Sickinger:varphiExp}. The dependence \eqref{Eq:CPR} of the CPR on the phase and applied field is the same as in our previous study\cite{Goldobin:2011:0-pi:H-tunable-CPR}. The difference is in the formulas \eqref{Eq:Gammas} for $\Gamma_0$ and $\Gamma_h$.

\fi


\bibliography{SF,pi}

\end{document}